\documentclass[twocolumn]{aastex631}

\usepackage{xspace}
\newcommand{\target}{\object{IIZw096}\xspace}
\newcommand{\LIR}{$L_{\rm{IR}}$\xspace}
\newcommand{\um}{$\mu$m\xspace}

\newcommand{\Nsrc}{12\xspace}

\begin{document}

\title{GOALS-JWST: Unveiling Dusty Compact Sources in the Merging Galaxy IIZw096}

\correspondingauthor{Hanae Inami}
\email{hanae@hiroshima-u.ac.jp}

\author[0000-0003-4268-0393]{Hanae Inami}
\affiliation{Hiroshima Astrophysical Science Center, Hiroshima University, 1-3-1 Kagamiyama, Higashi-Hiroshima, Hiroshima 739-8526, Japan}

\author[0000-0001-7291-0087]{Jason Surace}
\affiliation{IPAC, California Institute of Technology, 1200 E. California Blvd., Pasadena, CA 91125}

\author[0000-0003-3498-2973]{Lee Armus}
\affiliation{IPAC, California Institute of Technology, 1200 E. California Blvd., Pasadena, CA 91125}

\author[0000-0003-2638-1334]{Aaron S. Evans}
\affiliation{National Radio Astronomy Observatory, 520 Edgemont Road, Charlottesville, VA 22903, USA}
\affiliation{Department of Astronomy, University of Virginia, 530 McCormick Road, Charlottesville, VA 22903, USA}

\author[0000-0003-3917-6460]{Kirsten L. Larson}
\affiliation{AURA for the European Space Agency (ESA), Space Telescope Science Institute, 3700 San Martin Drive, Baltimore, MD 21218, USA}

\author[0000-0003-0057-8892]{Loreto Barcos-Munoz}
\affiliation{National Radio Astronomy Observatory, 520 Edgemont Road, Charlottesville, VA 22903, USA}
\affiliation{Department of Astronomy, University of Virginia, 530 McCormick Road, Charlottesville, VA 22903, USA}

\author{Sabrina Stierwalt}
\affiliation{Occidental College, Physics Department, 1600 Campus Road, Los Angeles, CA 90042}

\author[0000-0002-8204-8619]{Joseph M. Mazzarella}
\affiliation{IPAC, California Institute of Technology, 1200 E. California Blvd., Pasadena, CA 91125}

\author[0000-0003-3474-1125]{George C. Privon}
\affiliation{National Radio Astronomy Observatory, 520 Edgemont Road, Charlottesville, VA 22903, USA}
\affiliation{Department of Astronomy, University of Florida, P.O. Box 112055, Gainesville, FL 32611, USA}

\author[0000-0002-3139-3041]{Yiqing Song}
\affiliation{Department of Astronomy, University of Virginia, 530 McCormick Road, Charlottesville, VA 22903, USA}
\affiliation{National Radio Astronomy Observatory, 520 Edgemont Road, Charlottesville, VA 22903, USA}

\author[0000-0002-1000-6081]{Sean T. Linden}
\affiliation{Department of Astronomy, University of Massachusetts at Amherst, Amherst, MA 01003, USA}

\author[0000-0003-4073-3236]{Christopher C. Hayward}
\affiliation{Center for Computational Astrophysics, Flatiron Institute, 162 Fifth Avenue, New York, NY 10010, USA}

\author[0000-0002-5666-7782]{Torsten B\"oker}
\affiliation{European Space Agency, Space Telescope Science Institute, Baltimore, Maryland, USA}

\author[0000-0002-1912-0024]{Vivian U}
\affiliation{Department of Physics and Astronomy, 4129 Frederick Reines Hall, University of California, Irvine, CA 92697, USA}

\author[0000-0002-4375-254X]{Thomas Bohn}
\affiliation{Hiroshima Astrophysical Science Center, Hiroshima University, 1-3-1 Kagamiyama,  Higashi-Hiroshima, Hiroshima 739-8526, Japan}

\author[0000-0002-2688-1956]{Vassilis Charmandaris}
\affiliation{Department of Physics, University of Crete, Heraklion, 71003, Greece}
\affiliation{Institute of Astrophysics, Foundation for Research and Technology-Hellas (FORTH), Heraklion, 70013, Greece}
\affiliation{School of Sciences, European University Cyprus, Diogenes street, Engomi, 1516 Nicosia, Cyprus}

\author[0000-0003-0699-6083]{Tanio Diaz-Santos}
\affiliation{Institute of Astrophysics, Foundation for Research and Technology-Hellas (FORTH), Heraklion, 70013, Greece}
\affiliation{School of Sciences, European University Cyprus, Diogenes street, Engomi, 1516 Nicosia, Cyprus}

\author[0000-0001-6028-8059]{Justin H. Howell}
\affiliation{IPAC, California Institute of Technology, 1200 E. California Blvd., Pasadena, CA 91125}

\author[0000-0001-8490-6632]{Thomas Lai}
\affiliation{IPAC, California Institute of Technology, 1200 E. California Blvd., Pasadena, CA 91125}

\author[0000-0001-7421-2944]{Anne M. Medling}
\affiliation{Department of Physics \& Astronomy and Ritter Astrophysical Research Center, University of Toledo, Toledo, OH 43606, USA}
\affiliation{ARC Centre of Excellence for All Sky Astrophysics in 3 Dimensions (ASTRO 3D)}

\author[0000-0002-5807-5078]{Jeffrey A. Rich}
\affiliation{The Observatories of the Carnegie Institution for Science, 813 Santa Barbara Street, Pasadena, CA 91101}

\author[0000-0002-5828-7660]{Susanne Aalto}
\affiliation{Department of Space, Earth and Environment, Chalmers University of Technology, 412 96 Gothenburg, Sweden}

\author{Philip Appleton}
\affiliation{IPAC, California Institute of Technology, 1200 E. California Blvd., Pasadena, CA 91125}

\author[0000-0002-1207-9137]{Michael J. I. Brown}
\affiliation{School of Physics \& Astronomy, Monash University, Clayton, VIC 3800, Australia}

\author{Shunshi Hoshioka}
\affiliation{Hiroshima Astrophysical Science Center, Hiroshima University, 1-3-1 Kagamiyama,  Higashi-Hiroshima, Hiroshima 739-8526, Japan}

\author[0000-0002-4923-3281]{Kazushi Iwasawa}
\affiliation{Institut de Ci\`encies del Cosmos (ICCUB), Universitat de Barcelona (IEEC-UB), Mart\'i i Franqu\`es, 1, 08028 Barcelona, Spain}
\affiliation{ICREA, Pg. Llu\'is Companys 23, 08010 Barcelona, Spain}

\author[0000-0003-2743-8240]{Francisca Kemper}
\affiliation{Institut de Ciencies de l'Espai (ICE, CSIC), Can Magrans, s/n, 08193 Bellaterra, Barcelona, Spain}
\affiliation{ICREA, Pg. Llu\'s Companys 23, Barcelona, Spain}
\affiliation{Institut d'Estudis Espacials de Catalunya (IEEC), E-08034 Barcelona, Spain}

\author{David Law}
\affiliation{Space Telescope Science Institute, 3700 San Martin Drive, Baltimore, MD, 21218, USA}

\author[0000-0001-6919-1237]{Matthew A. Malkan}
\affiliation{Department of Physics \& Astronomy, UCLA, Los Angeles, CA 90095-1547}

\author{Jason Marshall}
\affiliation{Glendale Community College, 1500 N. Verdugo Rd., Glendale, CA 91208}

\author{Eric J. Murphy}
\affiliation{National Radio Astronomy Observatory, 520 Edgemont Road, Charlottesville, VA 22903, USA}

\author[0000-0002-1233-9998]{David Sanders}
\affiliation{Institute for Astronomy, University of Hawaii, 2680 Woodlawn Drive, Honolulu, HI 96822}

\author{Paul van der Werf}
\affiliation{Leiden Observatory, Leiden University, NL-2300 RA Leiden, Netherlands}



\begin{abstract} % 250 word limit 

We have used the Mid-InfraRed Instrument (MIRI) on the {\it James Webb Space Telescope} ({\it JWST}) to obtain the first spatially resolved, mid-infrared (mid-IR) images of \target, a merging luminous infrared galaxy (LIRG) at $z = 0.036$. Previous observations with the {\it Spitzer Space Telescope} suggested that the vast majority of the total IR luminosity (\LIR) of the system originated from a small region outside of the two merging nuclei. New observations with {\it JWST}/MIRI now allow an accurate measurement of the location and luminosity density of the source that is responsible for the bulk of the IR emission. We estimate that $40-70\%$ of the IR bolometric luminosity, or $3-5 \times 10^{11}\,{\rm{L_{\odot}}}$, arises from a source no larger than 175\,pc in radius, suggesting a luminosity density of at least $3-5 \times 10^{12} \, {\rm{L_{\odot} \, kpc^{-2}}}$. In addition, we detect 11 other star forming sources, five of which were previously unknown. The MIRI F1500W/F560W colors of most of these sources, including the source responsible for the bulk of the far-IR emission, are much redder than the nuclei of local LIRGs. These observations reveal the power of {\it JWST} to disentangle the complex regions at the hearts of merging, dusty galaxies.  

\end{abstract}




% ApJLetter limits
%   3500 words + 250 in abstract
%   5 figures or tables
%   50 references
% https://w.astro.berkeley.edu/~rdawson/countwords.html

\section{Introduction} \label{sec:intro}

\target (\object{CGCG448-020}, \object{IRAS20550+1656}) is a merging, luminous infrared galaxy (LIRG) at $z=0.0361$ with an infrared (IR) luminosity of $L_{\rm{IR,8-1000{\mu}m}}=8.7\times10^{11}L_\odot$, one of the more than 200 LIRGs in the Great Observatories All-sky LIRGs Survey \citep[GOALS;][]{ArmusL2009}. Previous imaging with the {\it Spitzer Space Telescope} revealed that the majority (up to $80\%$) of the infrared luminosity of the entire system comes from a region outside of the merging nuclei \citep{InamiH2010}, making it an even more extreme case than the well known Antennae Galaxies \citep{MirabelIF1998,BrandlBR2009}.

\begin{figure*}
  \includegraphics[width=\textwidth, clip, trim=0 0 0 0]{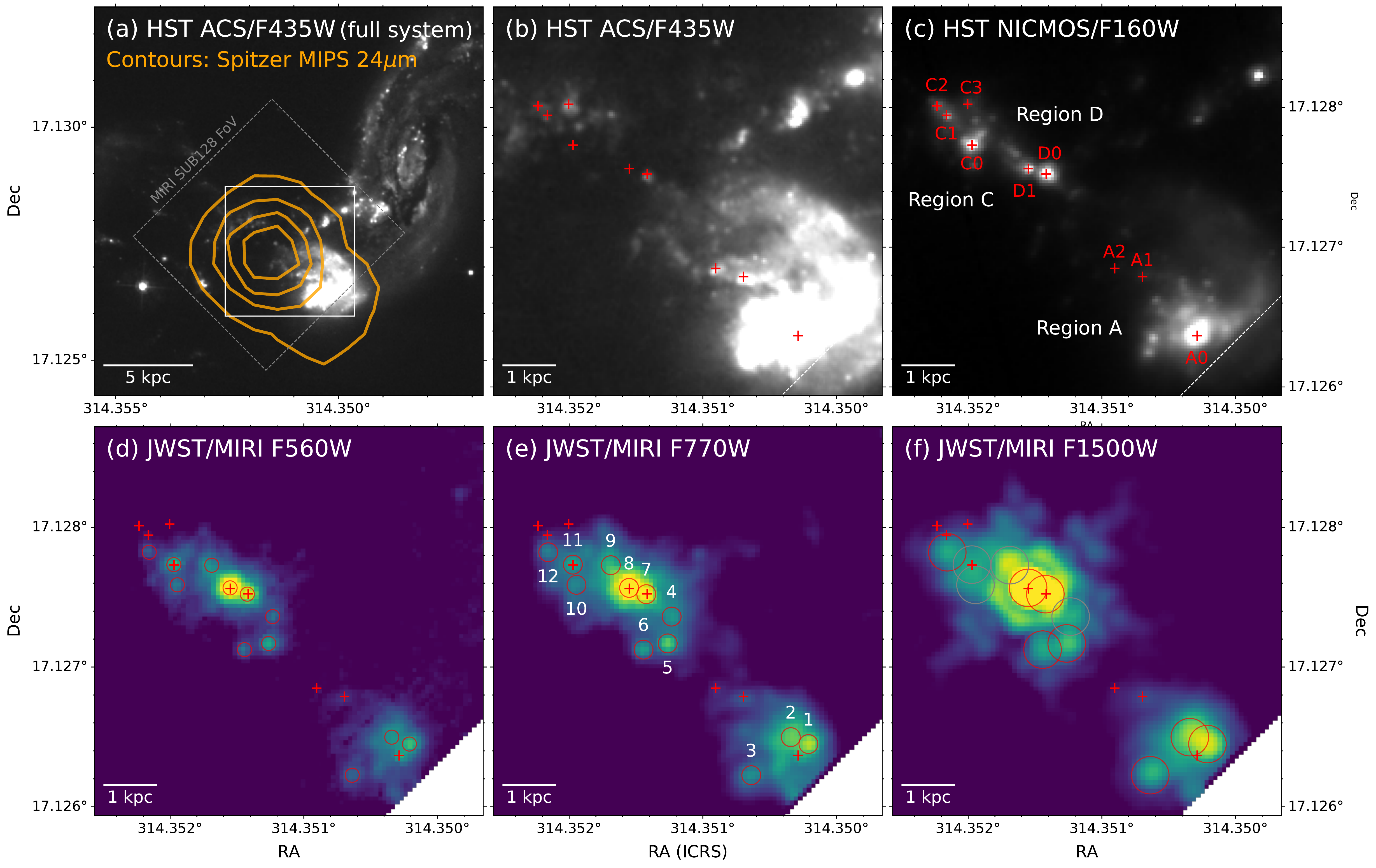}
  \caption{
  Multi-band imaging of \target.
  {\bf (a)} {\it HST}/ACS F435W (0.4\um) image showing the entire \target system with {\it Spitzer}/MIPS 24\um contours in orange. The FoV of the {\it JWST}/MIRI SUB128 subarray is centered on the dust-obscured region (gray dotted box). The white box indicates the region presented in the rest of the panels. {\bf (b)} The zoomed-in image of panel (a) showing the obscured region. The red plus symbols are a subset of the sources identified by \cite{WuH2022}. {\bf (c)} {\it HST}/NICMOS F160W (1.6\um) image of the obscured region. Region and source names from \cite{GoldaderJD1997} and \cite{WuH2022}, respectively, are shown. {\bf (d)--(f)} {\it JWST}/MIRI SUB128 images taken with the F560W (5.6\um), F770W (7.7\um), and F1500W (15\um) filters. The MIRI images are shown with a logarithmic scale. Red circles indicate the locations of the detected sources with a size corresponding to the beam FWHM.  Gray circles indicate sources detected in F560W and F770W but not confidently detected at F1500W. The red plus symbols are the sources detected with {\it HST} as shown in panels (b) and (c). PSF features are visible extending outwards from ID 8 in panel (f) because it is compact and bright. All images are shown with north up and east to the left. These images show that while the complexity of \target was evident from the near-IR {\it HST} data, the true nature of the dust emission and the source of the power is only finally revealed with {\it JWST}.
  \label{fig:MIRI_imgs}
  }
\end{figure*}

The system consists of regions A, C, and D (Figure~\ref{fig:MIRI_imgs}c) along with a merging spiral galaxy to the northwest \citep[][]{GoldaderJD1997}. Regions C and D are not detected or have extremely low signal-to-noise ratio (SNR) at ultraviolet and optical wavelengths with the {\it Hubble Space Telescope} ({\it HST}), but only at near-IR and longer wavelengths \citep{InamiH2010, Barcos-MunozL2017, WuH2022, SongY2022}. Although the {\it Spitzer}/MIPS 24\um image suggested that a single compact source in region D dominates the emission, the large beam size made it impossible to resolve the exact location of the immense far-IR emission. The superior sensitivity and resolving power of the {\it James Webb Space Telescope} ({\it JWST}) lets us pinpoint the source that is responsible for the intense IR emission and study its complex environment on sub-kpc scales in the mid-IR for the first time. 

Here, we present high-spatial resolution mid-IR imaging of \target taken with the {\it JWST} Mid-InfraRed Instrument \citep[MIRI;][]{RiekeGH2015, BouchetP2015}. Throughout this paper, we adopt a cosmology with $H=70 {\rm{km\,s^{-1}\,Mpc^{-1}}}$, $\Omega_{\rm{M}}=0.28$ and $\Omega_{\rm{\Lambda}}=0.72$. The redshift of \target ($z=0.0361$) corresponds to a luminosity distance of 160Mpc and a projected physical scale of 725${\rm{pc\,arcsec^{-1}}}$.

\section{Observations and Data Reduction} \label{sec:obs}

The {\it JWST} observations were performed under the Directors Discretionary Time Early Release Science (ERS) program 1328 (co-PIs: Armus, L. and Evans, A. S.). Images of \target were obtained on July 2, 2022 with a MIRI subarray (SUB128) using the F560W ($\lambda_{0}=5.6$\um), F770W (7.7\um) and F1500W (15\um) filters. The pointing was centered at 314.35167${\rm{^{\circ}}}$,17.12769${\rm{^{\circ}}}$ (J2000), where the prominent mid-IR emission was identified with {\it Spitzer}. The observations were dithered and exposure times set to avoid saturation (46, 48, and 48 seconds, respectively). The data were reduced with the standard {\it JWST} calibration pipeline \citep[data processing software ver.2022\_2a, calibration software ver.1.5.3;][]{GordonKD2015,BushouseH2022} and up-to-date reference files from the Calibration References Data System. The images in this work, including {\it HST}, are aligned to the {\it Gaia} Data Release 3 catalog \citep{GaiaCollaboration2016,GaiaCollaboration2021}.

\begin{figure}
  \includegraphics[width=0.48\textwidth, clip, trim=5 0 0 0]{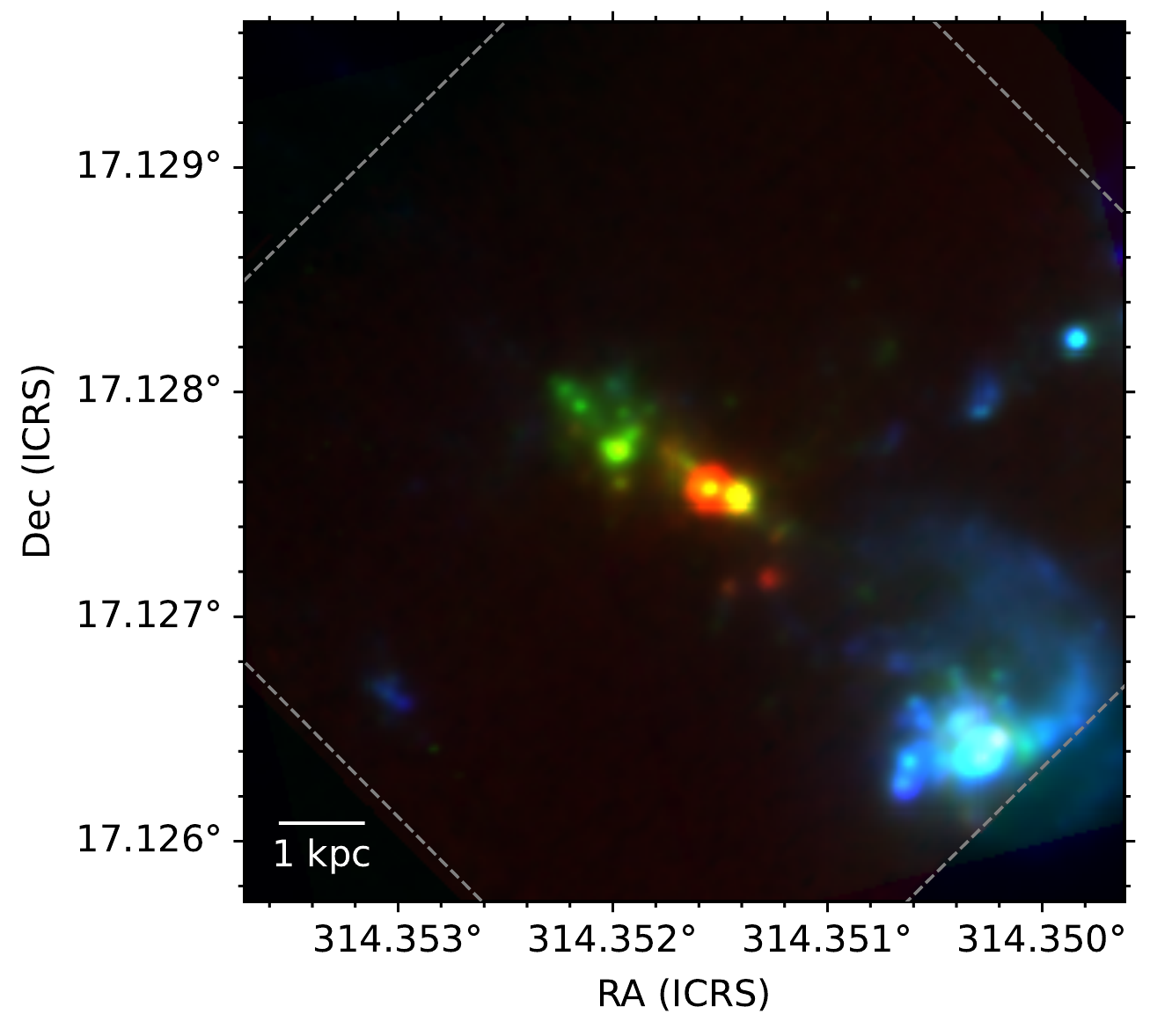}
  \caption{
  False color image of the \target obscured region, made with {\it JWST}/MIRI F560W (5.6\um, red), {\it HST}/NICMOS F160W (1.6\um, green), and {\it HST}/ACS F435W (0.4\um, blue). The displayed region is the same as in panels (b)-(f) of Figure~\ref{fig:MIRI_imgs}.
  \label{fig:RGB_imgs}
  }
\end{figure}

\section{Results} \label{sec:results}

The $\sim10\times$ improvement in the spatial resolution of {\it JWST} compared to {\it Spitzer} resolves the mid-IR emission into individual clumps down to scales of $\lesssim100-200$pc, enabling measurements of mid-IR color and \LIR surface density to study the nature of \target.

\subsection{Mid-IR Clumps in the Disturbed Region}

Individual sources are identified with the {\tt\string DAOFIND} algorithm \citep{StetsonPB1987} in the F770W SUB128 subarray image, which have the highest SNR, using the {\tt\string Python} {\tt\string photutils} package \citep{BradleyL2021}. The detection threshold is $5\sigma$. The sources detected in the F770W image are used as priors for photometry in all three bands (Figure~\ref{fig:MIRI_imgs}). We assign identification (ID) numbers for the detected sources in ascending order of R.A. For a subset of sources shown in \cite{WuH2022}, we also refer to their source names (Figure~\ref{fig:MIRI_imgs}c). The same \Nsrc clumps are detected in the F560W image, while in the F1500W image, four of them (IDs 4, 9, 10, 11/C0) lie on the structure of the point spread function (PSF) of the brightest source (ID 8/D1), making confident detections and flux measurements difficult. Thus, we only report their upper limits.

\begin{center}
\begin{table}
\centering
\caption{Flux density of the clumps detected at F560W, F770W, and F1500W \label{tbl:phot}}
\scalebox{0.75}{
\hspace*{-3.5cm}
\begin{tabular}{cccccc}
\hline
   ID &          RA &         Dec &                      F560W &                      F770W &                     F1500W \\ 
      &      degree &      degree &                        mJy &                        mJy &                        mJy \\ 
\hline
    1 &  314.350210 &   17.126449 & $  0.51^{+ 0.03}_{- 0.03}$ & $  3.17^{+ 0.16}_{- 0.23}$ & $  8.56^{+ 0.30}_{- 0.18}$ \\ 
    2 &  314.350342 &   17.126498 & $  0.35^{+ 0.03}_{- 0.05}$ & $  2.62^{+ 0.17}_{- 0.27}$ & $  7.10^{+ 0.36}_{- 0.25}$ \\ 
    3 &  314.350638 &   17.126226 & $  0.13^{+ 0.03}_{- 0.01}$ & $  0.87^{+ 0.06}_{- 0.08}$ & $  2.79^{+ 0.14}_{- 0.09}$ \\ 
    4 &  314.351233 &   17.127361 & $  0.13^{+ 0.03}_{- 0.02}$ & $  0.97^{+ 0.05}_{- 0.15}$ & $<            1.87$ \\ 
    5 &  314.351263 &   17.127169 & $  0.30^{+ 0.02}_{- 0.01}$ & $  1.60^{+ 0.05}_{- 0.06}$ & $  3.46^{+ 0.18}_{- 0.03}$ \\ 
    6 &  314.351444 &   17.127124 & $  0.13^{+ 0.03}_{- 0.01}$ & $  0.86^{+ 0.03}_{- 0.04}$ & $  2.27^{+ 0.13}_{- 0.04}$ \\ 
 7/D0 &  314.351423 &   17.127521 & $  1.79^{+ 0.05}_{- 0.05}$ & $  8.60^{+ 0.30}_{- 0.30}$ & $ 32.16^{+ 1.74}_{- 1.74}$ \\ 
 8/D1 &  314.351552 &   17.127566 & $  6.17^{+ 0.04}_{- 0.04}$ & $ 25.40^{+ 0.21}_{- 0.21}$ & $155.39^{+ 1.62}_{- 1.62}$ \\ 
    9 &  314.351688 &   17.127728 & $  0.31^{+ 0.04}_{- 0.03}$ & $  1.91^{+ 0.14}_{- 0.19}$ & $<            4.07$ \\ 
   10 &  314.351945 &   17.127588 & $  0.15^{+ 0.03}_{- 0.02}$ & $  0.75^{+ 0.09}_{- 0.18}$ & $<            2.80$ \\ 
11/C0 &  314.351972 &   17.127732 & $  0.39^{+ 0.02}_{- 0.03}$ & $  1.09^{+ 0.08}_{- 0.18}$ & $<            0.73$ \\ 
   12 &  314.352156 &   17.127821 & $  0.11^{+ 0.02}_{- 0.01}$ & $  0.67^{+ 0.04}_{- 0.05}$ & $  2.33^{+ 0.14}_{- 0.07}$ \\ 
\hline
Total &             &             & $ 10.45^{+ 0.36}_{- 0.33}$ & $ 48.52^{+ 1.39}_{- 1.94}$ & $214.06^{+ 4.63}_{- 4.04}$ \\ 
\hline
\end{tabular}
}
\end{table}
\end{center}

As shown in Figure~\ref{fig:MIRI_imgs}, the most prominent mid-IR source is ID 8 (source D1), lying $0.47^{\prime\prime}$ northeast of ID 7 (source D0). Although ID 8 is fainter than ID 7 in the {\it HST}/NICMOS 1.6\um image, its emission exceeds ID 7 at 5.6\um and 7.7\um by a factor of three. At 15\um, the ID\,8-to-ID\,7 flux ratio increases to about five. Although ID 7 was previously speculated to be associated with the bulk of the total IR emission due to its prominence at 1.6\um \citep{InamiH2010}, the majority of the mid-IR emission in fact originates from ID 8. 

There are five bright mid-IR clumps (IDs 1, 4, 5, 6, and 12) that are either not detected or have extremely low SNR in the {\it HST} 1.6\um image. ID 1 is in the less dusty region A to the southwest. The three sources, IDs 4, 5, and 6, are south of ID 7. The remaining one, ID 12, is located $2.27^{\prime\prime}$ northeast of ID 8. Interestingly, these new mid-IR selected sources are not concentrated in the dustiest region, but spread throughout the perturbed region.

Additional structure is evident in the MIRI image, outside of the main power source in the \target system (Figure~\ref{fig:RGB_imgs}). Region A, which accounts for most of the optical emission, hosts a number of clumps in the MIRI data. Around region C, the emission peaks at 1.6\um but fades towards the mid-IR.

\subsection{Mid-IR Colors of the Clumps}

Aperture photometry was employed to measure the flux of the detected clumps, except for IDs 7 and 8 due to their relative proximity (see below). The aperture radii used for F560W, F770W, and F1500W are $0.27^{\prime\prime}$, $0.28^{\prime\prime}$, and $0.30^{\prime\prime}$, respectively, with aperture corrections of 0.65, 0.65, and 0.50~\footnote{The aperture correction values are from the PSF encircled energy at
\url{https://jwst-docs.stsci.edu/jwst-mid-infrared-instrument/miri-performance/miri-point-spread-functions}}. Aperture photometry in the F1500W image was performed after subtracting ID 8, due to its prominent PSF pattern (note that the PSF has not been subtracted in Figure~\ref{fig:MIRI_imgs}f). The source subtraction was performed using a PSF generated by {\tt\string WebbPSF} \citep{PerrinMD2012,PerrinMD2014}~\footnote{\url{https://www.stsci.edu/jwst/science-planning/proposal-planning-toolbox/psf-simulation-tool}}. To account for the extremely red color of ID 8, a power-law spectrum with a spectral slope of 3, resembling the mid-IR color of \target, was used to generate the PSF. This provides a more accurate flux measurement of the sources around ID 8.

To extract the fluxes of IDs 7 and 8, a simultaneous two-dimensional Gaussian fit was performed in the F560W and F770W images. In the F1500W image, the flux of ID 8, which dominates the emission, was also extracted via a simultaneous two-dimensional Gaussian fit. However, we did not adopt the flux of ID 7 from this fit because this source lies on the Airy ring of ID 8. Instead, the flux of ID 7 was derived from a single Gaussian fit\footnote{We adopted Gaussian fits for IDs 7 and 8 to use a consistent method across the filter bands.} to the image after subtraction of ID 8 to minimize contamination from the PSF.

The local background was measured using a 3$\sigma$-clipped median of various annuli. They have a minimum inner radius $2.4\times$ the half-width half-maximum (HWHM) of the PSF and a maximum radius that is $14\times$ (for F560W and F770W) and $7\times$ (F1500W) the PSF HWHM in steps of $0.05^{\prime\prime}$ around each source. During this process, the other detected sources were masked out to avoid background over-estimation. Each of the measured background levels was separately subtracted from the measured flux, providing a distribution of fluxes for each source. The median value of this distribution is reported as the final flux density in Table~\ref{tbl:phot}. The 16th and 84th percentiles were adopted as the flux uncertainties. 

\begin{figure*}
  \includegraphics[width=\textwidth, clip, trim=0 0 0 0]{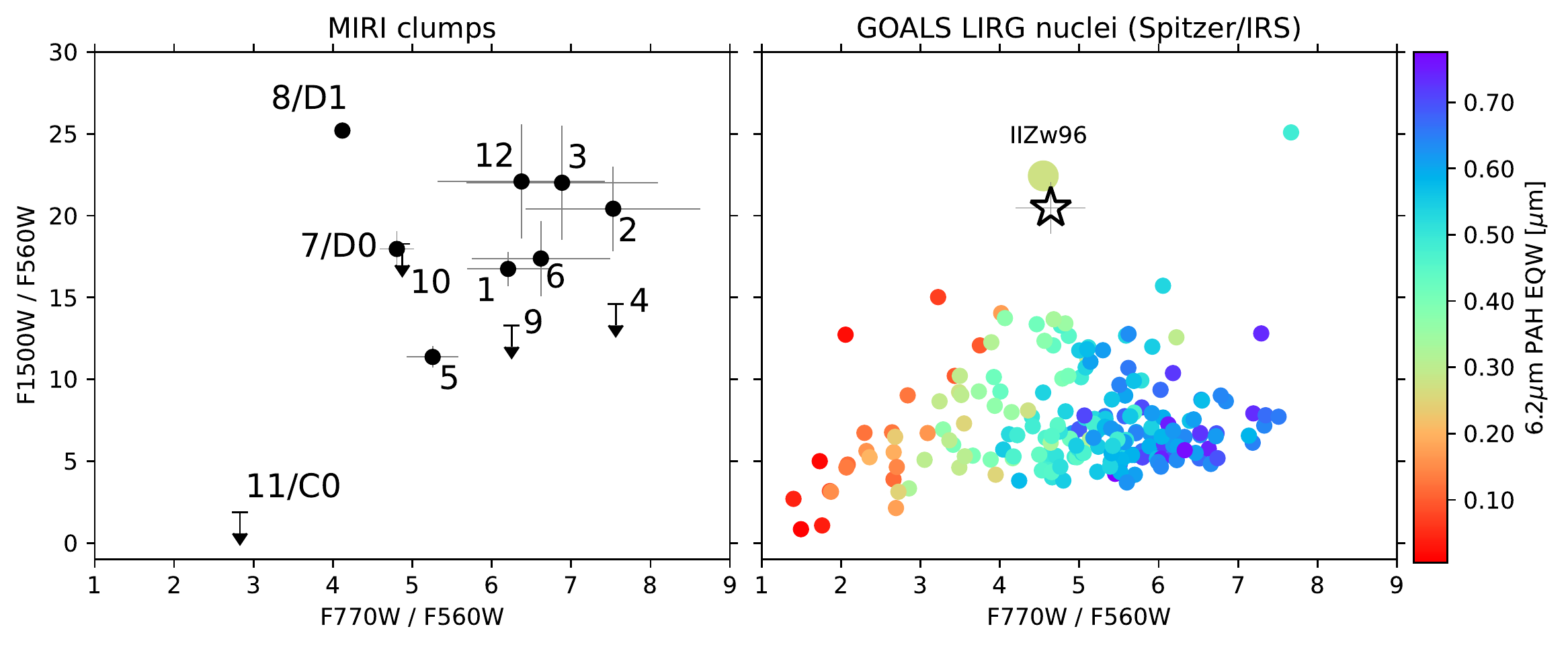}
  \caption{ 
    {\bf [Left]} F1500W/F560W$-$F770W/F560W color-color diagram for all clumps detected in the MIRI images. For sources without a detection in the F1500W image, upper limits are indicated by downward arrows. {\bf [Right]} For comparison, the same diagram but showing the colors of local LIRG nuclei derived from synthetic photometry on the {\it Spitzer}/IRS low-resolution spectra. The data points are color-coded by the 6.2\um PAH EQW measured from the spectra. The color derived from the total flux of the MIRI clumps in the left panel is shown as a star. In the {\it Spitzer} slit that covers the merger-induced dusty region, \target has one of the reddest F1500W/F560W colors among local LIRGs.
  \label{fig:CCD}}
\end{figure*}

Based on the measured fluxes in all three MIRI bands, we show the F1500W/F560W$-$F770W/F560W color-color diagram in Figure~\ref{fig:CCD}. These colors are a sensitive measure of the mid-IR continuum slope, and the F770W/F560W color is also sensitive to emission of polycyclic aromatic hydrocarbon (PAH) at 7.7\um. The individual clumps detected in the MIRI SUB128 images are shown in the left panel. We also present the same diagram for local LIRG nuclei using {\it Spitzer}/IRS low-resolution spectra taken with the Short-Low (SL; $5.5-14.5$\um) and Long-Low (LL; $14-38$\um) spectroscopy \citep{StierwaltS2013, StierwaltS2014}. We generated synthetic photometry from the spectra using the MIRI filter curves and the {\tt\string Python} {\tt\string synphot} package \citep{STScIDevelopmentTeam2018}. The SL and LL slits fully cover regions C and D. The LL slit also covers region A. As expected from the IRS spectra \citep[][]{InamiH2010,StierwaltS2013}, the F1500W/F560W color from the synthetic photometry of the dust-obscured region in \target is an outlier with a much redder color than the rest of the local LIRGs. With {\it JWST}/MIRI, we are now able to decompose the emission into individual clumps to study the distribution of their mid-IR colors.

The summed flux of all the clumps in color space agrees well with the color measured with the much larger beam of {\it Spitzer}/IRS (Figure~\ref{fig:CCD} right). All MIRI-detected clumps show redder colors (${\rm{F1500W/F560W}}\gtrsim15$) than most of the local LIRG nuclei, except for ID 5, which has F1500W/F560W$=11\pm0.7$. ID 8 (D1) is the reddest, while ID 7 (D0) has a comparable F1500W/F560W color to the other clumps. The newly detected source ID 12 is the second reddest, along with IDs 2 and 3 having excess emission at 15\um and 7.7\um compared to most of the other sources.

The F770W/F560W colors of the clumps are spread over a range of $4\lesssim{\rm{F770W/F560W}}\lesssim8$. As expected since they dominate the mid-IR flux, the F770W/F560W colors of the two brightest sources, IDs 7 and 8, agree well with the color of \target derived from the synthetic photometry on the IRS spectrum. The remaining sources, except for IDs 5, 10 and 11 are redder with ${\rm{F770W/F560W}}\gtrsim6$. These red sources have F770W/F560W colors consistent with strong PAH emission. The bluer sources suggest weaker PAH emission or an excess of hot dust. The F770W/F560W colors of IDs 7 and 8 are comparable to sources with 6.2\um PAH equivalent widths (EQWs) of about half that seen in pure starburst nuclei (Figure~\ref{fig:CCD} right), which is consistent with the direct measurement in the IRS spectrum \citep[6.2\um PAH EQW$=0.26$\um;][]{InamiH2010} and may indicate an excess of very hot dust. 

\subsection{Infrared Luminosity Surface Density}

To estimate the luminosity density of each clump in the MIRI image, we first use the measured flux to estimate \LIR. Assuming that the 15\um flux correlates with \LIR, we compute the fractional contribution of each clump to the total 15\um flux. The flux of the diffuse emission is measured with an aperture of radius $7^{\prime\prime}$ centered on the F1500W image with the total flux of all the clumps being subtracted. The resulting flux density of the diffuse emission is 190mJy~\footnote{In F560W and F770W, diffuse emission fluxes measured with the same method are 18mJy and 90mJy, respectively.}. Finally, the \LIR estimated in the obscured region \citep[$6.87\times10^{11}{\rm{L_\odot}}$;][]{InamiH2010} is scaled by these fractions to calculate the \LIR of each component. The resulting \LIR for ID 8 (D1) is $3\times10^{11}{\rm{L_\odot}}$, corresponding to a SFR of $40{\rm{M_{\odot}\,yr^{-1}}}$ \citep[assuming a Kroupa initial mass function;][]{KennicuttRCJ1998,KroupaP2001,MadauP2014}. This rises to $5\times10^{11}{\rm{L_{\odot}}}$ or a SFR of $60{\rm{M_{\odot}\,yr^{-1}}}$ if we assume that the diffuse emission at 15\um does not contribute to \LIR at all, because this ascribes more of the \LIR to ID 8. 

As an alternative, we derive a bolometric correction factor from the ensemble of GOALS nuclei to estimate \LIR of each clump. This factor is calculated using the 29 GOALS nuclei with similar colors to the clumps in \target, F1500W/F560W$>10$ and F770W/F560W$>4$ shown in Figure~\ref{fig:CCD} (right). From this we obtain a median bolometric correction factor (\LIR/$L_{\rm{\nu}}(15_{\rm{\mu m}})$) of $4\pm3$. This yields $L_{\rm{IR}}=(5\pm4)\times10^{11}{\rm{L_\odot}}$ for ID 8 (D1). This is consistent with the \LIR estimate above, where the total \LIR in the dust-embedded region is split up based on the 15\um flux fraction of each clump.

We can also estimate \LIR based on the known correlation between the 8\um luminosity ($L_{8}$) and \LIR \citep[$L_{\rm{IR}}/L_{8}=4.9^{+2.9}_{-2.2}$;][]{ElbazD2011}. However, $L_{8}$ for this correlation was obtained using the {\it Spitzer} 8\um band, which has a wider bandwidth than {\it JWST}/F770W (2.9\um and 2.2\um, respectively). Assuming that the bandwidth ratio can be used to correct for the difference, we obtain $L_{\rm{IR}}=(2\pm1)\times10^{11}{\rm{L_\odot}}$ for ID 8. This 8\um-based \LIR is slightly lower than, but consistent with the 15\um-based \LIR. Because the longer IR wavelength better traces the IR bolometric luminosity, we adopt the 15\um-based \LIR hereafter.

Because ID 8 (D1) is unresolved, the MIRI PSF size at 15\um limits its radius to $<175$pc. Thus, a lower limit on the \LIR surface density (${\Sigma}L_{\rm{IR}}$) is $>3-5\times10^{12}{\rm{L_{\odot}\,kpc^{-2}}}$. This corresponds to a SFR surface density of $>400-600{\rm{M_{\odot}\,yr^{-1}\,kpc^{-2}}}$ if powered by star formation.

\section{Discussion} \label{sec:discussion}

Our {\it JWST} imaging has revealed the complexity of the dustiest region of the merging galaxy, \target. The three main components in the perturbed region show a wide variety of optical-infrared colors and morphologies, with a mix of bright, unresolved clumps and diffuse emission. This suggests a range in properties, such as extinction, SFR, age, and dust temperature in this ongoing merger.

Although ID 7 (D0) is the brightest source at 1.6\um, the {\it JWST} mid-IR data demonstrate that ID 8 (D1) generates the bulk of the total IR emission in \target. The location of ID 8 also coincides with two OH megamasers \citep{MigenesV2011,WuH2022}~\footnote{\target also hosts an $\rm{H_{2}O}$ megamaser \citep{WigginsBK2016,KuoCY2018}, but its exact location is unknown.}. Megamasers are often found in merging (U)LIRGs, in close proximity to the nuclei \citep[e.g.,][]{RobertsH2021}, marking regions of extremely high gas density and strong far-IR radiation.

The emission we have targeted with {\it JWST} is clearly responsible for the bulk of the luminosity in \target and it arises from outside of the prominent two merging galaxies, one of which is the spiral galaxy to the northwest that lies outside of the MIRI SUB128 FoV (Figure~\ref{fig:MIRI_imgs}a). However, it is possible that ID 8 is a third nucleus in this system. Given the observed mid-IR morphology, region A$+$C$+$D could be a single disrupted galaxy or it could be two galaxies with source A0 being one nucleus and ID 8 (D1) being the other. The diffuse, extended emission around ID 8 would then be the remnants of the third galaxy's disk. The estimated stellar mass of ID 8 is $\sim10^{9}{\rm{M_{\odot}}}$ and there is a similarly large mass of gas in this region  \citep{InamiH2010,WuH2022}. These estimates might indicate ID 8 is a partially stripped third nucleus. Although the current MIRI SUB128 images do not provide evidence for either the two- or three-body merger scenario, our upcoming {\it JWST} observations (ERS program 1328) may elucidate this question. The deeper MIRI and NIRCam full-array imaging may detect a more pronounced disk-like morphology around ID 8. A detection of a rotation curve around ID 8 by the planned spectroscopic observations could suggest that ID 8 is a third nucleus.

ID 8 (D1) generates $40-70\%$ of the total IR emission of the \target system, corresponding to an \LIR surface density of $>3-5\times10^{12}{\rm{L_{\odot}\,kpc^{-2}}}$. This is $\sim10\times$ the characteristic surface brightness of starbursts, but comparable to super star clusters \citep{MeurerGR1997} including some in the Antennae Galaxies \citep[][]{BrandlBR2009}. The \LIR surface density limit of ID 8 is also consistent with the ULIRG nuclei studied at 12.5\um with Keck \citep[][]{SoiferBT2000,SoiferBT2001}. In addition, the 33GHz continuum imaging ($0.1^{\prime\prime}$ resolution) taken with the Very Large Array shows that the peak emission is located at ID 8 \citep{SongY2022}. These authors estimated a SFR surface density of $470\pm60{\rm{M_\odot\,yr^{-1}\,kpc^{-2}}}$, corresponding to an \LIR surface density of $(3.9\pm0.5)\times10^{12}{\rm{L_{\odot}\,kpc^{-2}}}$, which agrees with the value derived from the mid-IR. Given the \LIR surface density limit from the mid-IR and the column density of $\sim10^{25}{\rm{cm^{-2}}}$ obtained via the molecular gas mass \citep[within an aperture of $0.2^{\prime\prime}\times0.16^{\prime\prime}$;][]{WuH2022}, ID 8 appears to be below the Eddington limit if its size is $175$\,pc. \citep[e.g., ][]{Pereira-SantaellaM2021,Barcos-MunozL2015}.

The clumps in the disturbed region, including ID 8 (D1), are much redder in F1500W/F560W than local LIRG nuclei. Based on the 9.7\um silicate optical depth ($\tau_{9.7{\rm{\mu m}}}\sim1$) derived from {\it Spitzer}/IRS spectroscopy, the $V$-band extinction is estimated to be $\geq19$mag \citep{InamiH2010}. The unusually red colors of \target could be due to an extremely young, highly obscured starburst or AGN, triggered by the recent merger. The other LIRG with a very red F1500W/F560W color, IRAS22491-1808, is also an on-going merger with bright clumps of star formation \citep{SuraceJA1998a,SuraceJA1998}. 

The F770W/F560W color traces the 7.7\um PAH emission and is also a good proxy for the 6.2\um PAH EQW (Figure~\ref{fig:CCD} right). Using the 6.2\um PAH EQW as a diagnostic of starbursts and AGN \citep[e.g.,][]{BrandlBR2006,ArmusL2007,PetricAO2011}, the clumps with F770W/F560W$\gtrsim5$ are consistent with pure star formation. However, the color of ID 8 (D1) is in the range where an AGN cannot be excluded. Although analysis of X-ray spectra of \target obtained with {\it Chandra}, {\it XMM-Newton}, and {\it NuSTAR} also favor star formation, the non-detection of ID 8 by {\it NuSTAR} does not rule out a Compton-thick AGN if the column density exceeds $10^{25}{\rm{cm^{-2}}}$ \citep{IwasawaK2011,RicciC2021}. In fact, the estimated column density of ID 8 is $\sim10^{25}{\rm{cm^{-2}}}$ \citep[][]{WuH2022}. The evidence may be consistent with the presence of a Compton-thick AGN but it is equally consistent with a starburst, and neither is conclusive.
Our upcoming mid-IR and near-IR spectroscopic data (ERS program 1328) are expected to shed additional light on the underlying energy source of this heavily obscured source, perhaps through detection of one or more coronal emission lines.

\section{Conclusions} \label{sec:conclusion}

The {\it JWST}/MIRI imaging demonstrates uncharted aspects of the dust emission from the extremely luminous, merging galaxy \target. The high spatial resolution and high sensitivity mid-IR imaging of this work yields the following findings:

 \begin{itemize}

     \item For the first time, we have spatially resolved the mid-IR emission of the merger-induced heavily dust-obscured region of \target. We identify the source (ID 8/D1) that is responsible for the bulk of the mid-IR emission, accounting for $40-70$\% of the total IR emission of the system. 
     
     \item In total, 12 clumps are detected in the F770W (and F560W) image, five of which are newly identified and were not detected or had low SNR detections at 1.6\um with {\it HST}/NICMOS. Most of the clumps have similar F1500W/F560W colors, ranging from $\sim15$ to 25. These colors are about twice as red as local LIRG nuclei, but agree with the colors derived from synthetic photometry on {\it Spitzer} spectra of this system. Among LIRG nuclei, the F770W/F560W colors roughly correlate with the 6.2\um PAH EQW, and therefore the clumps have colors indicative of 6.2\um PAH EQWs from $\sim0.3$\um to $0.6$\um, slightly lower than but including pure star formation. 

     \item The estimated \LIR of ID 8 (D1) is $3-5\times10^{11}{\rm{L_{\odot}}}$, which corresponds to a SFR of $40-60{\rm{M_{\odot}\,yr^{-1}}}$ if it is star-forming. As the source is unresolved, we estimate its \LIR surface density to be $>3-5\times10^{12}{\rm{L_{\odot}\,kpc^{-2}}}$ or a SFR surface density of $>400-600{\rm{M_{\odot}\,yr^{-1}\,kpc^{-2}}}$. Such high surface densities put source D1 in a range comparable to young super star clusters and ULIRG nuclei. 

 \end{itemize}

The {\it JWST} mid-IR imaging described in this {\it Letter} has revealed a hidden aspect of \target, and has opened a door towards identifying heavily dust-obscured sources which cannot be found at shorter wavelengths. Future planned spectroscopic observations of \target will provide additional information on the nature of the dust, ionized gas, and warm molecular gas in and around the disturbed region of this luminous merging galaxy.

%% Also note that the akcnowlodgment environment does not support long amounts of text. If you have a lot of people and institutions to acknowledge, do not use this command. Instead, create a new \section{Acknowledgments}.
\begin{acknowledgments}
The authors would like to thank the referee whose constructive comments helped improve the manuscript.
% MAST DOI
The {\it JWST} data presented in this paper were obtained from the Mikulski Archive for Space Telescopes (MAST) at the Space Telescope Science Institute. The specific observations analyzed can be accessed via \dataset[https://doi.org/10.17909/8c47-wb74]{https://doi.org/10.17909/8c47-wb74}. 
STScI is operated by the Association of Universities for Research in Astronomy, Inc., under NASA contract NAS5–26555. Support to MAST for these data is provided by the NASA Office of Space Science via grant NAG5–7584 and by other grants and contracts.
% JWST
This work is based on observations made with the NASA/ESA/CSA James Webb Space Telescope.
These observations are associated with program 1328.
% MAST
The data were obtained from the Mikulski Archive for Space Telescopes at the Space Telescope Science Institute, which is operated by the Association of Universities for Research in Astronomy, Inc., under NASA contract NAS 5-03127 for JWST.
% NED
This research has made use of the NASA/IPAC Extragalactic Database (NED),
which is funded by the National Aeronautics and Space Administration and
operated by the California Institute of Technology.
% Hanae Inami
HI acknowledges support from JSPS KAKENHI Grant Number JP19K23462 and the Ito Foundation for Promotion of Science.
% Yiqing Song
YS is supported by the NSF through grant AST 1816838 and the Grote Reber Fellowship Program administered by the Associated Universities, Inc./ National Radio Astronomy Observatory.
% Christopher C. Hayward
The Flatiron Institute is supported by the Simons Foundation.
% Vivian U
VU acknowledges funding support from NASA Astrophysics Data Analysis Program (ADAP) grant 80NSSC20K0450. 
% Anne M. Medling
AMM acknowledges support from the National Science Foundation under Grant No. 2009416.
% Susanne Aalto
S.A. gratefully acknowledges support from an ERC Advanced
Grant 789410, from the Swedish Research Council and from the Knut and Alice Wallenberg (KAW) foundation.
% Kazushi Iwasawa
KI acknowledges support by the Spanish MCIN under grant PID2019-105510GB-C33/AEI/10.13039/501100011033.
\end{acknowledgments}


\vspace{5mm}
\facilities{
  JWST (MIRI), 
  HST (ACS, NICMOS)
  NED
           }


\software{
  astropy \citep{AstropyCollaboration2013,AstropyCollaboration2018},  
  photutils \citep{BradleyL2021},
  synphot \citep{STScIDevelopmentTeam2018},
  WebbPSF \citep{PerrinMD2012,PerrinMD2014}
          }



% \appendix

% \section{Appendix information}


\bibliography{references.bib}{}
\bibliographystyle{aasjournal}



%@arxiver{IIZw096_HST_MIRI_SUB128_images_6panels_noRGB.pdf,IIZw096_MIRI_SUB128_RGB_F560W_F160W_F435W.pdf,IIZw096_MIRI_SUB128_CCD_final_tbl.pdf}

\end{document}